\documentclass[reprint,superscriptaddress,showpacs,amsfonts,amsmath,amssymb,prl]{revtex4-1}
\usepackage{graphicx}
\usepackage{hyperref}
\usepackage{textcomp}
\begin{document}

\title{Observation of anomalous Hanle spin precession lineshapes resulting from \\ interaction with localized states}

\author{J.~J.~van~den~Berg}
\email{j.j.van.den.berg@rug.nl}
\author{B.~J.~van~Wees}
\affiliation{Physics of Nanodevices, Zernike Institute for Advanced Materials, University of Groningen, Nijenborgh 4, 9747 AG Groningen, The Netherlands}
\author{W.~Strupinski}
\affiliation{Institute of Electronic Materials Technology, Wolczynska 133, 01-919 Warsaw, Poland}

\begin{abstract}
It has been shown recently that in spin precession experiments, the interaction of spins with localized states can change the response to a magnetic field, leading to a modified, effective spin relaxation time and precession frequency. Here, we show that also the shape of the Hanle curve can change, so that it cannot be fitted with the solutions of the conventional Bloch equation. We present experimental data that shows such an effect arising at low temperatures in epitaxial graphene on silicon carbide with localized states in the carbon buffer layer. We compare the strength of the effect between materials with different growth methods, epitaxial growth by sublimation and by chemical vapor deposition. The presented analysis gives information about the density of localized states and their coupling to the graphene states, which is inaccessible by charge transport measurements and can be applied to any spin transport channel that is coupled to localized states.
\end{abstract}


\maketitle



The spin relaxation length $\lambda$ is a property that can be obtained in a (non-local) spin valve geometry by two independent methods. Firstly, by measuring the spin dependent signal as a function of the distance $x$ to the spin injecting electrode, because this signal scales with the spin accumulation $\mu_S\propto e^{-x/\lambda}$. The spin relaxation time $\tau$ is then given by $\tau=\lambda^2/D_S$, where $D_S$ is the spin diffusion coefficient that can be obtained from charge transport measurements. A complementary method is Hanle spin precession, where an out-of-plane field $B_z$ is applied to precess the spins in the x-y plane of the channel with precession frequency $\omega_L$. $D_S$ and $\tau$ are directly obtained by numerically fitting the Hanle curve with the stationary solutions of the one-dimensional Bloch equation:
\begin{equation}
	\label{eq:bloch}
	\vec{0}=D_S\nabla^2\vec{\mu_S}-\frac{\vec{\mu_S}}{\tau}+\vec{\omega_L}\times \vec{\mu_S}.
\end{equation}
Here, $\vec{\omega_L}=g\mu_B\vec{B}/\hbar$, with the Land\'{e} g-factor g=2, $\mu_B$ the Bohr magneton, and $\hbar$ the reduced Planck's constant. 

\begin{figure}[h] 
    \includegraphics[width=\columnwidth]{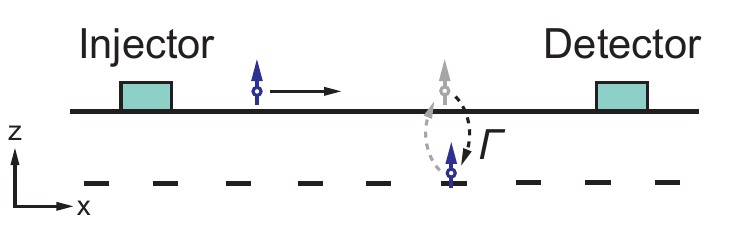}
    \caption{A spin transport channel coupled to localized spin states with a coupling rate $\Gamma$. Spins can reside in the localized states for some time, where they pick up extra relaxation and precession. This has a strong effect on the measured Hanle signal at the detector.}
    \label{fig:0}
\end{figure}

A striking effect occurs when itinerant electron spins couple to localized spin states. The injected spins pick up extra relaxation and precession while residing for some time in the localized states on their way from injector to detector, as depicted in Fig.~\ref{fig:0}. This causes a significant decrease in the measured spin signal amplitude and a narrowing of the Hanle curve. The effect was observed in Epitaxial Graphene (EG) on Silicon Carbide (SiC),\cite{maassen_long_2012} a system that receives much interest as a platform for spin transport experiments,\cite{maassen_long_2012,dlubak_highly_2012,birkner_annealing-induced_2013} because of its high quality, large area, insulating substrate and more recently, experimental findings of room temperature ferromagnetism.\cite{xie_room_2011,giesbers_interface-induced_2013} The localized states most likely originate from the bufferlayer, which is the nonconducting graphene-like layer between channel and substrate.\cite{maassen_localized_2013} Though firstly developed for observations in EG on SiC, the localized states picture is in fact applicable to any spin transport channel which couples to localized states. Recently, Roundy and Raikh\cite{roundy_spin_2014} theoretically studied the effect of deep traps in organic semiconductors and predicted a narrowing of the Hanle curve in a similar fashion. By making assumptions on the distribution of the dwell times in the traps, an anomalous shape of the Hanle curve was also predicted, but not yet confirmed by experiment.

In this paper we show an anomalous lineshape of the Hanle curve in spin precession experiments in EG at low temperatures. Also, we present analysis that can qualitatively explain the observed anomalous behavior by assuming a transition regime of strong to weak coupling between the channel and the localized states. The effect is measured in EG on SiC samples that we produced by two growth methods: by sublimation \cite{berger_ultrathin_2004,virojanadara_homogeneous_2008,emtsev_towards_2009,strupinski_growth_2009} and by CVD (Chemical Vapor Deposition).\cite{al-temimy_low_2009,strupinski_graphene_2011} Raman spectra have shown a structural difference between the buffer layers of the two materials.\cite{strupinski_graphene_2011} Therefore, a difference in coupling and/or density of localized states can be expected, resulting in a measurable effect on the spin relaxation time and the precession frequency in the channel. Our analysis allows for estimating the density of localized states and the spin relaxation time for both growth methods. 



We investigated two different types of EG spin valve devices: EG by sublimation (Type I) and on EG by CVD (Type II). We show here the results of one device of each type, similar measurements on another type II device can be found in the supplementary information.

For both types,  we grew the graphene layers at 1600~\textdegree C under an argon (Ar) laminar flow in a hot-wall Aixtron VP508 reactor on semi-insulating on-axis oriented 4H-SiC(0001) substrates with dimensions 5~mm $\times$ 5~mm. The dynamic flow conditions in the reactor controlled the Si sublimation rate. For epitaxial CVD growth, we formed an Ar boundary layer thick enough to prevent Si sublimation, but allowing for the diffusion of propane gas that was led into the reactor as the precursor. The reactor pressure applied in the case of sublimation was 100~mbar and in the case of CVD 30~mbar.\cite{strupinski_growth_2009,strupinski_graphene_2011}. These two growth methods resulted in a wafer scaled EG monolayer, with a nonconducting carbon buffer layer sandwiched between the graphene and the SiC. Raman studies showed that the buffer layer of Type II has less epitaxial strain, probably due to a different superstructure or C--C bond length. This results in slightly higher mobility than in Type I, but comparable doping.\cite{strupinski_graphene_2011} 

\begin{figure}[h] 
    \includegraphics[width=\columnwidth]{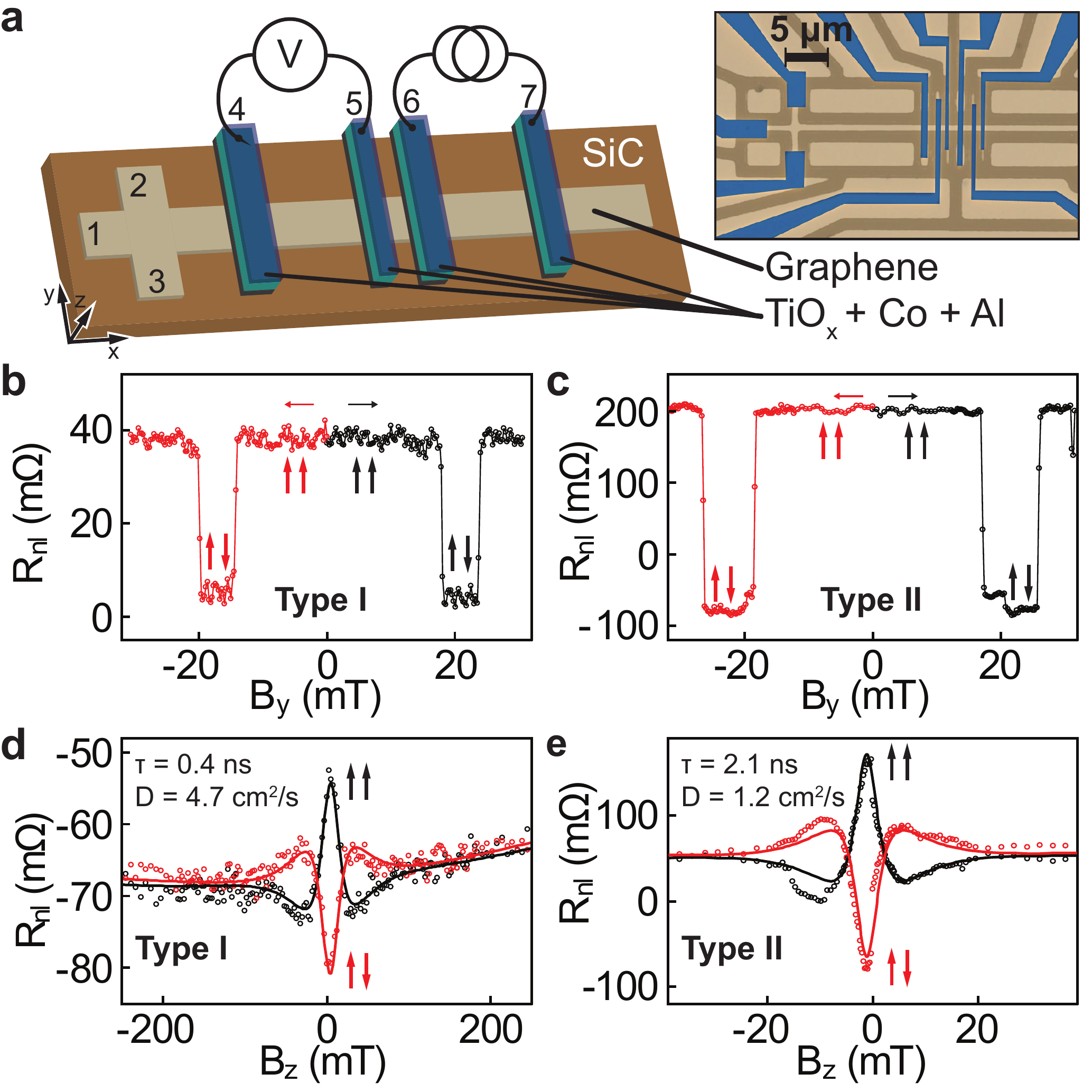}
    \caption{(a) Device schematics of an EG on SiC nonlocal spin valve with four spin contacts made of a TiO$_2$ tunnel barrier, Co and an Al capping layer. The inset shows a colored microscopic image of the real device (Type I) with a microscopic image of the etch lines (dark brown) superimposed. Additional lines were etched away in between the Co contacts to ensure there were no current paths outside the strip. (b)$\--$(c) RT spin valve measurements of EG by sublimation (Type I) and CVD (Type II). The distance $L$ between the inner electrodes is 1~\textmu m for both devices. The horizontal arrows represent the sweeping direction of the magnetic field. Vertical arrows show parallel or antiparallel orientation of the inner electrodes. The extra level in (c) is caused by the switching of one of the outer electrodes. The lines between measurement points are a guide to the eye. (d)$\--$(e) Hanle precession measurements at RT of type I and II in the parallel (black circles) and antiparallel (red circles) configuration. Note the large difference in the x-axis scales. The best fits are performed on a plot of $(R_{nl}^{\uparrow\uparrow}-R_{nl}^{\uparrow\downarrow})/2$ and inserted in this figure as black and red lines. Because of a small asymmetry in (e) the fit is performed only in the positive range of $B_z$.}
    \label{fig:1}
\end{figure}

We produced EG spin transport devices with ferromagnetic contacts using standard lithography techniques, following roughly the same recipe as described in Ref.~\onlinecite{maassen_long_2012}.
We used EBL (Electron Beam Lithography) to define an EG strip with a Hall cross at one end of the strip as shown in Fig.~\ref{fig:1}(a). \footnote{We avoided the use of any negative tone resist because this can leave problematic resist residue after exposure, leading to high contacts resistances and low polarization.} For this, we wrote $\sim$0.5$\--$1.0~\textmu m wide lines to define the strip. We used reactive ion etching in an RF O$_2$ plasma for 45 seconds (at 25~W and 0.01~mbar) for removal of the EG. In the final EBL step we defined the contact pattern. Using e-beam evaporation we deposited 0.9~nm of titanium oxide (TiO$_2$) in two steps, which we oxidized in situ in O$_2$ atmosphere after each step at a pressure above 10$^{-1}$~mbar. The TiO$_2$ acts as the tunnel barrier to avoid conductivity mismatch. \cite{schmidt_fundamental_2000,maassen_contact-induced_2012} Without breaking the vacuum we then deposited 40~nm of cobalt (Co) and a 5~nm aluminium (Al) capping layer to prevent the Co from oxidizing. Figure~\ref{fig:1}(a) shows a schematic representation as well as a colored microscopic image of this device. The devices were loaded in a liquid helium flow cryostat and measured in vacuum ($\sim$10$^{-7}$~mbar). 

We performed room temperature Hall measurements to characterize our nanodevices. We used contacts 1--4 in Fig.~\ref{fig:1}(a) to obtain $n$. \footnote{We note that in our device geometry the Hall measurement probes a small EG region close to, but not exactly at the spin channel.} We found the samples to be n-doped with a charge carrier density of $n_1$ = 1.5 $\times$ 10$^{12}$ and $n_2$ = 1.2 $\times$ 10$^{12}$~cm$^{-2}$ and a mobility of $\mu_1$ = 2800 and $\mu_2$ = 3400~cm$^2$V$^{-1}$s$^{-1}$. \footnote{Epitaxial graphene on SiC is known to be heavily n-doped. Hall characterization of the full 5 $\times$ 5~mm graphene sample yielded values of 2--6 $\times$ 10$^{12}$~cm$^{-2}$} These values correspond to a charge diffusion coefficient $D_{C1}=200$ and $D_{C2}=172$~cm$^2$s$^{-1}$. \footnote{We obtained the charge diffusion coefficient $D_C$ using the Einstein relation $D_C=1/(R_{sq}e^2\nu(E_F))=\hbar v_F/(2R_{sq}e^2(g_sg_v\pi n)^{1/2})$, where $R_{sq}$ is the square resistance, $e$ the electron charge, $\nu(E_F)$ the density of states at the Fermi level, the valley and spin degeneracy $g_v$ and $g_s$ equal to 2, the Fermi velocity $v_F\approx10^6$~m~s$^{-1}$, and $n$ the charge carrier density.}

To investigate the spin transport properties we used the nonlocal spin valve geometry shown in Fig.~\ref{fig:1}(a). Using standard lock-in techniques, we sent an AC current $I_{AC}$ (= 1$\--$5~\textmu A) between Co contacts 6 and 7, creating a nonequilibrium spin accumulation in the channel underneath these contacts. The spin accumulation decays in both positive and negative x-direction and can be detected as a nonlocal voltage $V_{nl}$ between spin sensitive contacts 4 and 5. A spin valve measurement consists of the nonlocal resistance $R_{nl}=V_{nl}/I_{AC}$ as a function of an in-plane field $B_y$. Because of a different width of each contact and thereby a difference in their coercive fields, it is possible to switch their magnetization direction independently. A sudden change in $R_{nl}$ can be observed when the relative orientation of the injector and detector changes from parallel to antiparallel or vice versa. When the outer contacts are far away at a distance greater than $\lambda$, we can neglect their influence, resulting in a typical two-level spin valve.

Figures~\ref{fig:1}(b)$\--$(c) show room temperature (RT) spin valve measurements of Type I ($L$ = 1~\textmu m and $W$ = 0.85~\textmu m) and Type II ($L$ = 1~\textmu m and $W$ = 0.62~\textmu m). We measured spin signals $(R_{nl}^{\uparrow\uparrow}-R_{nl}^{\uparrow\downarrow})$ of 35~m$\Omega$ and 280~m$\Omega$ respectively. The spin valve signal in Fig.~\ref{fig:1}(c) shows a third level in $R_{nl}$, which is caused by the small contribution of one of the outer contacts.

Figures~\ref{fig:1}(d)$\--$(e) show the results of the Hanle precession experiments that we performed on Type I and Type II EG devices at RT. Both devices show very narrow Hanle curves when compared to typical exfoliated graphene, as was already reported for Type I devices in Ref.~\onlinecite{maassen_long_2012} (for similar measurements on a second Type II device, see supplementary). Using Eq.~\ref{eq:bloch}, we obtained a high $\tau$ of 400 $\pm$ 30~ps and 2.1 $\pm$ 0.09~ns, as well as a very low $D_S$ of 4.7 $\pm$ 0.4 and 1.2 $\pm$ 0.1~cm$^2$s$^{-1}$ for Type I and Type II respectively. The clearly observable zero crossing and pronounced shoulders ensure a reliable fitting and therefore an accurate value of $D_S$ in case of a graphene channel without coupling.\cite{maassen_contact-induced_2012} However, the obtained value for $D_S$ is highly underestimated in the case of a graphene channel coupled to localized states as described in the model of Ref.~\onlinecite{maassen_localized_2013}. This is underlined by the significant deviation from the diffusion coefficient we measured in the charge transport experiment, $D_{C1}\approx43D_{S1}$ and $D_{C2}\approx143D_{S2}$.

Next, we look at the transport model introduced in Ref.~\onlinecite{maassen_localized_2013}. This model assumes electronic states coupled to the channel that do not contribute to the charge transport. However, the states can contribute to the spin relaxation and precession. This happens when the coupling is strong enough and there are enough states contributing, a condition that can be described by the expression $\eta\Gamma\gg1/\tau'$. Here, $\Gamma$ is the coupling rate between the channel and the localized states, $\eta=\nu_{ls}/\nu_{graphene}$ is the ratio between density of states in the graphene and the density of localized states and $\tau'$ is the intrinsic spin relaxation time of the graphene, unaffected by the influence of the localized states. The dynamics in the system are described by an effective Bloch equation:
\begin{equation}
\label{eq:effbloch}
\vec{0}=D_S'\nabla^2\vec{\mu_S}-\frac{\vec{\mu_s}}{\tau^{eff}}+\vec{\omega_L^{eff}}\times \vec{\mu_S}.
\end{equation}
We denote $D_S'$ to indicate that this is the diffusion coefficient of graphene unaffected by the localized states. The effective relaxation time $\tau^{eff}$ and effective precession frequency $\omega_L^{eff}$ are given by:
\begin{equation}
\label{eq:efftau}
\frac{1}{\tau^{eff}}=\frac{1}{\tau'}+\eta\Gamma\frac{1+\tau^*\Gamma+(\tau^*\omega_L^*)^2}{(1+\tau^*\Gamma)^2+(\tau^*\omega_L^*)^2}
\end{equation}
and
\begin{equation}
\label{eq:effomega}
\omega_L^{eff}=\omega_L'+\eta\Gamma^2\frac{(\tau^*)^2\omega_L^*}{(1+\tau^*\Gamma)^2+(\tau^*\omega_L^*)^2}.
\end{equation}
where $\omega_L'$ is the precession frequency in the graphene unaffected by the localized states. In the above expressions a $*$ is assigned to the corresponding properties of the localized states.

The parameter space in this description includes seven quantities, $D_S'$, $\tau'$, $\tau^*$, $g$, $g^*$, $\Gamma$, $\eta$, but we reduce that number by making some assumptions. First, we assume the properties of the graphene channel to be the same as the typical properties of exfoliated graphene with similar mobility: $D_S'=200$~cm$^2$s$^{-1}$, $\tau'=150$~ps and $g=2$. \cite{tombros_electronic_2007} We can make this assumption, because 1) we measured $D_C\approx D_S'$ and 2) due to weak spin-orbit interaction and the absence of electron-electron interaction we have no reason to expect a change in $g$.  We can estimate $\eta=D_S'/D_S-1\approx D_C/D_S$, under the reasonable assumptions that 1) $g=g^*=2$ and 2) the room temperature data is in the strong coupling limit. Under the same assumption we can estimate $\tau^*$, using the expression  $1/\tau=1/\tau'+\eta/\tau^*$. \footnote{In this notation $D_S$ and $\tau$, the values we obtained by fitting with Eq.~\ref{eq:bloch}, are the same as $D_S^{mod}$ and $\tau^{mod}$ from Ref.~\onlinecite{maassen_localized_2013}} The only unknown variable we are now left with is the coupling rate $\Gamma$. 



\begin{figure}[h]
    \includegraphics[width=\columnwidth]{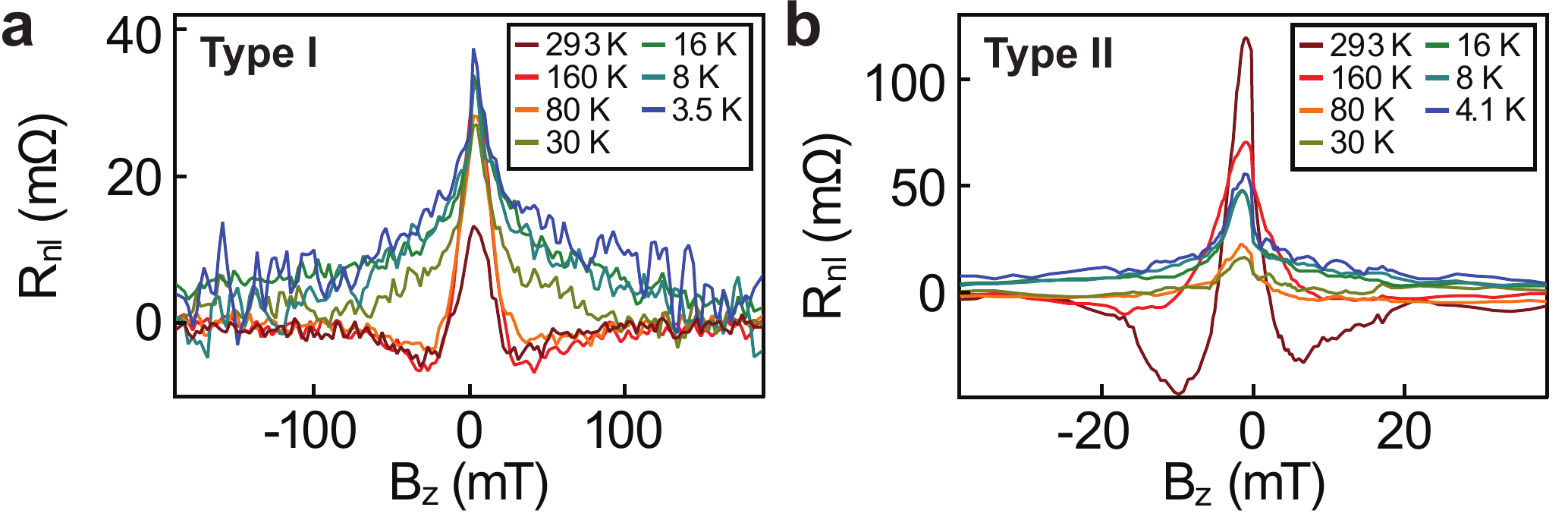}
    \caption{Temperature dependent evolution of the Hanle curve. (a) EG by sublimation (Type I). The typical RT Hanle lineshape (dark red line) develops into a broader curve without zero-crossings in $R_{nl}$ and with a narrow peak around zero magnetic field at temperatures of 30~K and lower. (b) EG by CVD (Type II). The Hanle lineshapes change already at 160~K and also the amplitude decreases in the temperature range 293$\--$30~K and increases again at 30$\--$4.1~K.}
    \label{fig:2}
\end{figure}

An interesting question related to the nature of the coupling to the localized states, is whether the process is mediated by temperature independent tunneling or hopping, which should be temperature dependent. Therefore, we cooled down the samples to liquid helium temperatures. We observed a distinct change in the lineshape of the Hanle curve when cooling down. Figures~\ref{fig:2}(a)$\--$(b) show the temperature dependence of the Hanle curve for both types. We plot here the spin signal $(R_{nl}^{\uparrow\uparrow}-R_{nl}^{\uparrow\downarrow})/2$ for each individual curve. At temperatures of 30~K and lower, the Hanle curve of Type I in Fig~\ref{fig:2}(a) evolves into a broader curve that does not cross zero and has an extra, narrow peak around zero field. This temperature dependence of the Hanle curve is surprising, as it was not seen before in EG by sublimation.\cite{maassen_long_2012,birkner_annealing-induced_2013} Type II in Fig~\ref{fig:2}(b) behaves comparably, but also shows a decrease in amplitude at lower temperatures and the transition already takes place around 160~K.

Now we show in what regime of parameters these expressions can result in a Hanle curve with an anomalous shape. According to Ref.~\onlinecite{maassen_localized_2013} $\tau^{eff}$ is approximately constant and $\omega_L^{eff}$ linear with magnetic field in both the weak coupling regime ($\eta\Gamma\ll1/\tau'$) and the strong coupling regime ($\eta\Gamma\gg1/\tau'$). In both limits the Hanle curve has the same shape, but in the case of strong coupling the amplitude and peak width are reduced drastically. This explains the modified transport properties and the difference between $D_C$ and $D_S$ that we typically see in EG at RT. Only in the transition regime ($\eta\Gamma\sim1/\tau'$) the full expression of Eq.~\ref{eq:efftau} and Eq.~\ref{eq:effomega} comes into play. A nontrivial lineshape then arises from the fact that $\tau^{eff}$ becomes a function of $\omega_L$ and thus scales with $B_z$.

\begin{figure}[h]
    \includegraphics[width=\columnwidth]{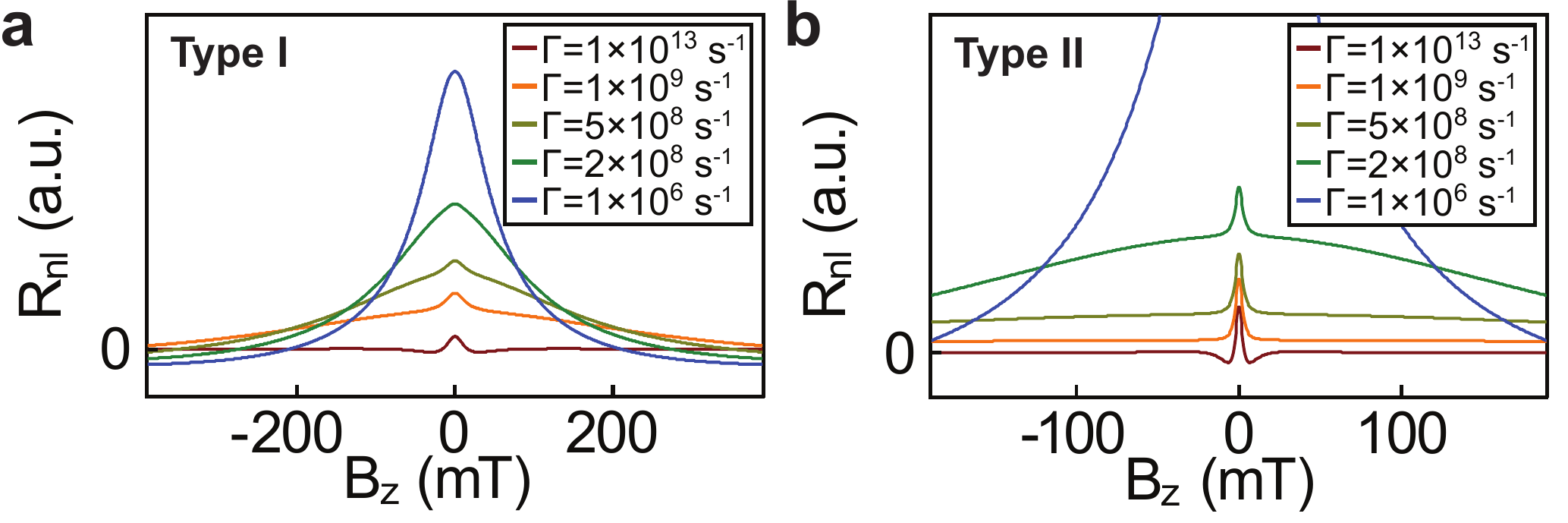}
    \caption{Simulated curves for Hanle precession including the effect of localized states. The figures show the evolution of the Hanle lineshape when changing the coupling rate $\Gamma$. For both plots we used $D_S'$ = 200~cm$^2$s$^{-1}$, $\tau'$ = 150~ps and $g$ = 2. The blue curves show the weak coupling limit which is not influenced by the properties of the localized states and is therefore the same in both figures. (a) uses the quantities $\tau^*$ and $\eta$ of Type I and b) uses $\tau^*$ and $\eta$ of Type II.}
    \label{fig:3}
\end{figure}

In Fig.~\ref{fig:3} we model the development of the Hanle curve when entering the transition regime between strong and weak coupling. For these plots we use the solution to Eq.~\ref{eq:effbloch} and fill in Eq.~\ref{eq:efftau} and Eq.~\ref{eq:effomega}. We use the experimental results for $D_S$, $\tau$ [Fig.~\ref{fig:1}(d)$\--$(e)] and $D_C$ to estimate $\eta=42$ (166) and $\tau^*=0.42$~(2.3)~ns for Type I (II). In the weak coupling regime (blue curve) both Hanle curves follow the spin properties of the channel. In the model we assume these  properties to be equal for both devices, so the blue curves in Fig~\ref{fig:3}(a) and (b) are the same. In the intermediate regime, which is around $20$~ps $\leq1/\eta\Gamma\leq200$~ps, the typical Hanle lineshape changes into a superposition of a broad and a narrow peak. In the strong coupling regime (dark red lines) the shape is the same as a regular Hanle curve, but its amplitude and peak width both decrease drastically. In this regime the spin properties are dominated by the localized spin states, therefore $\tau^*\approx\tau$.

We now discuss how the simulated data from the localized states model relates to the experiment. When we compare Fig.~\ref{fig:2} and \ref{fig:3} we see a similar double peak shape emerging when going from high to low temperature or from strong to weak coupling, respectively. The similarity of the two figures points to a coupling rate that is temperature dependent. This can be explained by two possible mechanisms: 1) the localized states are at the Fermi energy and the coupling is mediated by activated transport, where the thermal energy is used to overcome a tunnel barrier or 2) the localized states are distributed over a range of energies, depending on the available thermal energy. The strength of the effect is different for the two materials under investigation, EG by sublimation and EG by CVD. This can be explained by the structural difference of the bufferlayer, resulting in a different density and\slash or energy distribution of the localized states.

When comparing the experiment and simulations we see a difference in ratio between the signal amplitude and curve width. This suggests more of the fitting parameters are temperature dependent, limiting the usefulness of Eq.~\ref{eq:effbloch} for fitting the experimental data. We also attempted to simulate the data using more free fit parameters. This, however, leads to a result with too many uncertainties of which the interpretation is unclear and an even higher value for $\eta$ (see Supplementary Information for the details of this analysis). A possibility would be to incorporate a distribution of coupling rates similar to a so-called heavy tail distribution of trapping times that is seen in disordered semiconductors,\cite{tiedje_physical_1981} which influences the Hanle shape even outside the intermediate coupling regime according to Ref.~\onlinecite{roundy_spin_2014}. Such an analysis might also shed more light on the high value we find for $\eta$, giving $\nu_{ls}=\eta\nu_{graphene}\approx2\times10^{13}$~eV$^{-1}$cm$^{-2}$ or higher. This, however, is outside the scope of this paper.



In short, we investigated a new regime of intermediate coupling strength between diffusive and localized electron spins in EG on SiC and observed anomalous Hanle curves, giving novel insights in the effects that can play a role in widely used Hanle precession experiments. The interaction between channel and localized states is complex, reflected in the number of variables in the model. Nonetheless, we found a qualitative agreement between the temperature dependence of the anomalous experimental Hanle curves and our model. We observed a significant difference in the density of localized states between the two material types, which is likely to be related to the structure of the bufferlayer, the effect being stronger in EG by CVD (Type II). For both types, the analysis leads to a high density of localized states in the buffer layer.

On the one hand this work might provide new answers for how to control transported spins by coupling them to their environment. On the other it confirms the usefulness of spin transport to ``read out'' local spin properties.



\begin{acknowledgements}
We would like to acknowledge H. M. de Roosz, H. Adema and J. G. Holstein for technical support and J. Fabian and P.J. Zomer for feedback on the manuscript. The research leading to these results has received funding from NanoNed, the Zernike Institute for Advanced Materials and the European Union Seventh Framework Programmes under Grant Agreement "ConceptGraphene" (\textnumero257829) and "Graphene Flagship" (\textnumero604391).
\end{acknowledgements}




%

\newpage

\section{\label{sec:Sup}Supplementary information}

\begin{figure}[h]
	\includegraphics[width=0.48\textwidth]{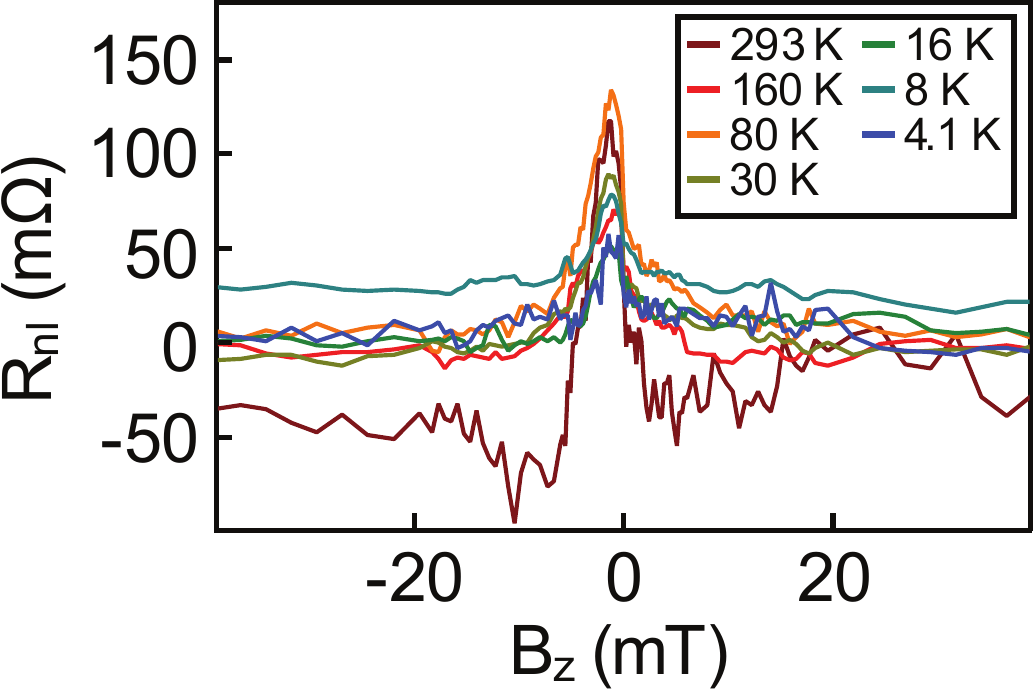}
	\caption{Temperature dependent evolution of the Hanle curve of another EG by CVD (Type II) sample. The signal is more unstable and noisy than in Fig.~3(b) of the main text and the amplitude of the signal is fluctuating between subsequent measurements. However, the change from a regular Hanle lineshape at RT into an anomalous shape at lowered temperatures can still be seen.}
	\label{fig:s1}
\end{figure}

\begin{figure}[h]
	\includegraphics[width=0.48\textwidth]{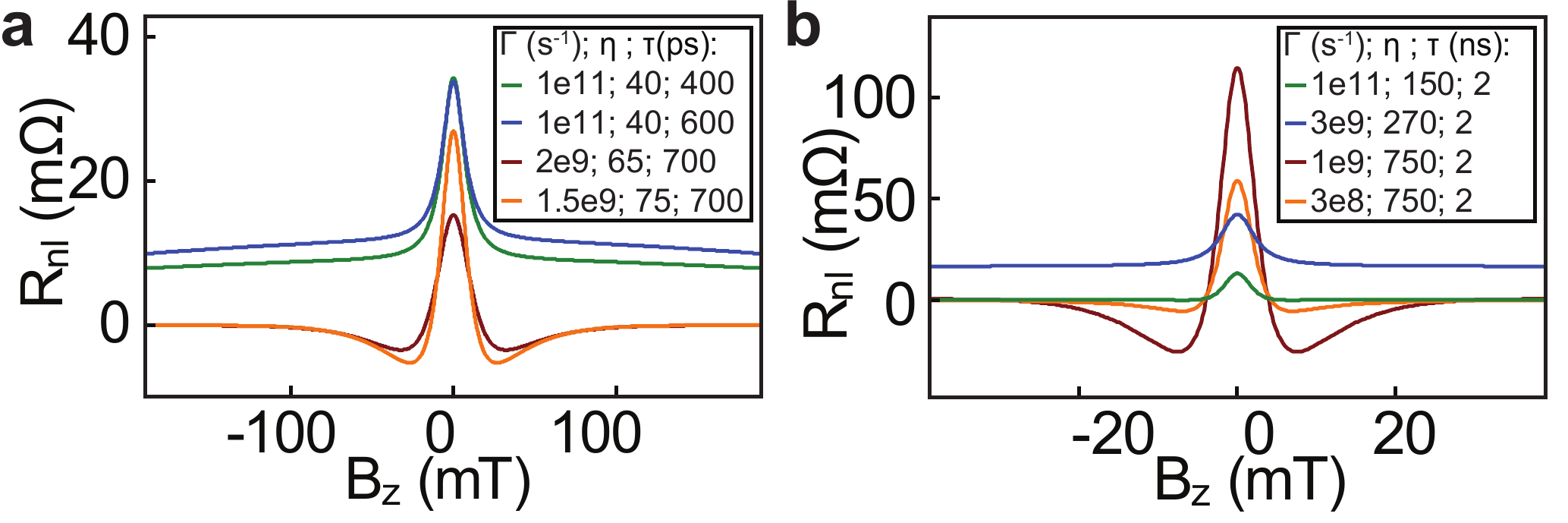}
	\caption{Attempt to simulate the temperature dependent, experimental data shown in Fig.~3 of the main text by assuming $\Gamma$, $\eta$ and $\tau^*$, to be temperature dependent. It is possible to obtain a reasonable agreement, but the fits are very sensitive to small changes in the fitting parameters.}
	\label{fig:s2}
\end{figure}

Figure~\ref{fig:s1} is a temperature dependent Hanle precession measurement of a second type II sample, showing a similar change in the lineshape as in Fig.~3(b) of the main text but with a higher noise level. Also, the amplitude of the signal fluctuates between subsequent measurements. Fitting this data with the model would require the parameters $\eta$ and $\Gamma$ to have a nontrivial dependence on temperature, showing that our model is limited to only qualitatively explaining the effect of anomalous Hanle lineshapes.

The information from the temperature dependent measurements can be used to estimate the spin properties of both the channel and the localized states, as well as the coupling $\Gamma$ and the ratio $\eta$. Here, we try to reproduce the experimental Hanle curves from Fig.~3 of the main text using more free fit parameters. In Fig~\ref{fig:s2}(a) we compare the model to the data from Type I. We assume three quantities, $\Gamma$, $\eta$ and $\tau^*$, to be temperature dependent for a reasonable match with the experimental data. Surprisingly, for a reasonable agreement we need an increase in $\eta$, meaning that the density of localized states contributing to the spin transport increases with temperature. The same trend can be seen for Type 2 [Fig.~\ref{fig:s2}(b)], although this requires high values of $\eta$ = 750.

\end{document}